\documentclass[twocolumn,eqsecnum,aps,prl]{revtex4}

\usepackage[dvips]{graphicx}
\usepackage{dcolumn}
\usepackage{bm}

\begin{document}

\preprint{APS/123-QED}

\title{Specific Heat Study of an ${\bm S}=$1/2 Alternating Heisenberg Chain System F$_5$PNN \\ In Magnetic Field }
\author{Y. Yoshida, 
N. Tateiwa,
M. Mito,
T. Kawae, K. Takeda, Y. Hosokoshi,$^1$$^{\dagger }$ and K. Inoue$^{1}$$^{\ddagger }$
}
\address{Department of Applied Quantum Physics, Faculty of Engineering, Kyushu University, Fukuoka 812-8581, Japan \\
$^{1}$ Institute for Molecular Science, Nishigounaka 38, Myodaiji, Okazaki, Aichi 444-8585, Japan}

\date{\today}

\begin{abstract}
We have measured 
the 
specific heat of an 
$S=1/2$ antiferromagnetic alternating Heisenberg chain
pentafulorophenyl nitronyl nitroxide in magnetic fields up to $H>H_\mathrm{C2}$. 
This compound has the field-induced magnetic ordered (FIMO) phase between 
$H_\mathrm{C1}$ and $H_\mathrm{C2}$. 
Characteristic behaviors are observed depending on the magnetic field up to above $H_\mathrm{C2}$ 
outside of the $H$-$T$ boundary for the FIMO. 
Temperature and field dependence of the specific heat 
are qualitatively in good agreement with the theoretical calculation on 
an $S=1/2$ two-leg ladder. 
[Wang {\it et al.} Phys. Rev. Lett {\bf 84} 5399 (2000)] 
This agreement suggests that the observed behaviors are related with the low-energy excitation in the Tomonaga-Luttinger liquid.

\end{abstract}
\maketitle

Since the Haldane conjecture \cite{Haldane}, 
the quasi-one-dimensional (1D) antiferromagnetic spin-gapped systems, 
have been attracting much attention from both experimental and theoretical points of view. 
Spin-gapped systems, such as Haldane systems, spin-Peierls systems, two-leg ladder systems and alternating chain systems 
have a finite energy gap $\Delta$ between the non-magnetic ground state and the lowest excited state 
because of notable quantum effects. 
When magnetic field is applied to these systems, 
the lowest branch of the first excited state goes down due to the Zeeman effect and 
intersects the ground state at the lower critical field $H_\mathrm{C1}$=$\Delta/g \mu_\mathrm{B}$, 
indicating that the gap closes.
For the higher field, the magnetization starts to grow and saturates at the upper critical field $H_\mathrm{C2}$ with a spin-polarized state.
For $H_\mathrm{C1}\leq H\leq H_\mathrm{C2}$, 
the system has the gapless excitation on the magnetic ground state,
and is described by the universarity class of 1D quantum systems called the Tomonaga-Luttinger liquid (TLL), 
in which heuristic features, such as the power-law temperature dependence of the NMR relaxation rate and 
the linear-$T$ dependence of the low-temperature specific heat are theoretically predicted \cite{TLL}.
Experimentally, Chaboussant {\it et al.} reported the observation of the TLL state in 
an $S=1/2$ two-leg ladder Cu$_2$(C$_5$H$_{12}$N$_2$)Cl$_4$ (CuHpCl) in the NMR study in magnetic fields up to above $H_\mathrm{C2}$ \cite{chabou}. 
As for the previous specific heat measurements 
on the spin-gapped compounds, however, most of those results 
were limited far below $H_\mathrm{C2}$ 
and the probing of the TLL behavior by measuring the specific heat 
in the entire field range of the gapless phase has not been done [4, 5, 6-8].

The compound we focus on here is a genuine organic radical crystal pentafulorophenyl nitronyl nitroxide (F$_5$PNN).
The magnetism of F$_5$PNN comes from an unpaired electron 
localized 
around NO groups in the molecule.
The low-temperature susceptibility is well reproduced 
by an alternating chain model 
with alternating exchange interactions along the chain, $J_{1}/k_\mathrm{B}=-2.8$ K, $J_{2}/J_{1}=0.4$ \cite{Hosokoshi}.
This compound has two-fold advantages. 
Firstly, it provides us with a really isotropic quantum spin system because of the quenching of the spin-orbital interactions.
In other words, we are free from the anisotropy effects which are ordinarily inevitable in inorganic compounds. 
Secondly, this compound has an appropriate values of $H_\mathrm{C1}$ and $H_\mathrm{C2}$ for the experimental 
study.
From the low-temperature magnetization process, 
$H_\mathrm{C1}$ and $H_\mathrm{C2}$ were estimated to be 3 T and 6.5 T, respectively \cite{Hosokoshi}. 
It should be noted here that these two critical fields 
are much smaller than those of other spin-gapped compounds 
and can be easily attained with a conventional superconducting magnet, 
which enables us to perform the detailed specific heat measurements on F$_5$PNN 
in the three different phases.
\begin{figure}[bp]
\includegraphics*[width=5.5cm]{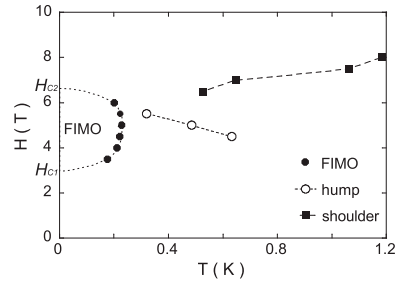}
\caption{\label{fig:PD}  Magnetic field versus temperature phase diagram for F$_5$PNN based on data from the previous \cite{ore} and present specific heat measurements. }
\end{figure}

In our previous paper, 
we have found the field-induced magnetic ordering (FIMO) in F$_5$PNN by the specific heat measurements 
in magnetic fields up to 6 T and at around 200 mK \cite{ore}. 
At $H=5$ T, a characteristic hump was observed at around 500 mK outside of the FIMO phase in the $H$-$T$ phase diagram. 
On the other hand,  Izumi {\it et al.} reported that the NMR relaxation rate $T_1 ^{-1}$ of F$_5$PNN 
showed an anomalous power-law behavior below about 4 K for $H_\mathrm{C1}<H<H_\mathrm{C2}$ \cite{izumi}. 
From the experimental point of view, however, we have to be careful about the effect of the inevitable inter-chain coupling $J'$ 
on the present physics of interest. 
In F$_5$PNN, $J'/k_\mathrm{B}$ is estimated to be $1\times 10^{-2}$ K by the mean-field theory for FIMO in the gapless phase, 
where the three-dimensional ordering temperature $T_\mathrm{C}$ is given as  $T_\mathrm{C}\approx \frac{\sqrt{J'J}}{k_\mathrm{B}}$ 
(Here we take ${J}/{k_\mathrm{B}}$ as the average value of ${J_1}/{k_\mathrm{B}}$ and ${J_2}/{k_B}$ and 
$T_\mathrm{C}\simeq$ 200 mK at $H=5$ T for an order estimation). 
The value of $J'/k_\mathrm{B}$ gives an exchange field $H_\mathrm{ex}\simeq \frac{2J'\langle S\rangle}{g\mu _\mathrm{B}}$ 
of about $10^2$ Oe, about 0.5 \% for the external field $H=5$ T. 
Thus, the hump in the specific heat observed outside of the FIMO boundary is intrinsic to the one-dimensionality of F$_5$PNN, and 
should be studied more in detail. 

In this letter, 
we extend the specific heat measurement in magnetic field up to $H=8$ T ($>$ $H_\mathrm{C2}$) and in the temperature range up to 6 K. 
In addition to the sharp anomaly due to FIMO, a hump or a shoulder in the specific heat is observed in the field range above $H_\mathrm{C1}$. 
The temperatures of these anomalies are plotted in $H$-$T$ phase diagram of Fig. 1, together with our previous results \cite{ore}. 
Although the low-temperature specific heat at the critical field $H\simeq H_\mathrm{C1}$ shows linear-$T$ dependence, 
it does not seem to approach zero monotonously as  $T$ $\rightarrow$ 0 K at $H=2.5$ T and 7 T, where the system is just outside of the gapless phase. 
These results will be discussed in connection with the TLL behaviors which were reported in the above NMR study 
and in the theoretical specific heat study of the 
two-leg ladder system \cite{wang}. 

The sample preparation of F$_5$PNN is detailed in Ref. \cite{Hosokoshi2}. 
Specific heat measurements on the polycrystalline sample of F$_5$PNN were performed by the adiabatic heat-pulse method using a $^3$He-$^4$He dilution refrigerator.

Figure 2 shows some representative data for temperature dependence of the total specific heat $C$ in the magnetic fields up to 8 T. 
First, we give characteristic behaviors of $C$ in some typical fields.
At $H=0$ T, $C$ shows a broad maximum around $T=2$ K, which is due to the short-range ordering in a 1D magnetic system. 
At low temperatures, $C$ decreases exponentially with decreasing temperature, 
reflecting the energy gap above the non-magnetic ground state. 
As magnetic field increases, 
the broad maximum is suppressed and broadened. 
In the field $H=3$ T($\simeq  H_\mathrm{C1}$), $C$ shows linear-$T$ dependence. 
At 5 T, a sharp anomaly due to the FIMO is seen at $T\simeq  200$ mK and 
the hump appears at around 500 mK with its maximum. 
These results are consistent with our previous work. 
In magnetic fields higher than $H_\mathrm{C2}$($\simeq  6.5$ T), 
$C$ shows a shoulder in the temperature range of 500 mK$<T<1.5$ K 
and decreases exponentially with decreasing temperature. 
In order to see these features more clearly, we estimated temperature dependence of the magnetic specific heat $C_\mathrm{m}$ 
by subtracting  the lattice contribution $C_\mathrm{l}$ from $C$.
A broken line in Fig. 2 shows $C_\mathrm{l}$ 
which is estimated from the data at $H=0$ T so that the total magnetic entropy for $N$ spins should approach $Nk_\mathrm{B}\mathrm{ln}(2S+1)$ 
at higher temperatures. 
$C_\mathrm{m}$ will be shown in the following figures 
in the four field regions covering the two critical fields 
$H_\mathrm{C1}$ and $H_\mathrm{C2}$.
\begin{figure}[t]
\includegraphics*[width=5.5cm]{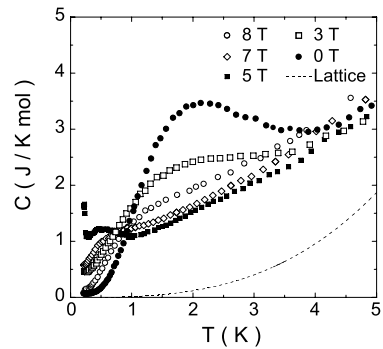}
\caption{\label{fig:Ctotal}  Some representative data for temperature dependence of $C$ of F$_5$PNN in magnetic fields up to 8 T. 
A broken line shows the lattice contribution.
}
\end{figure}


\begin{figure*}[t]
\includegraphics*[width=17.8cm]{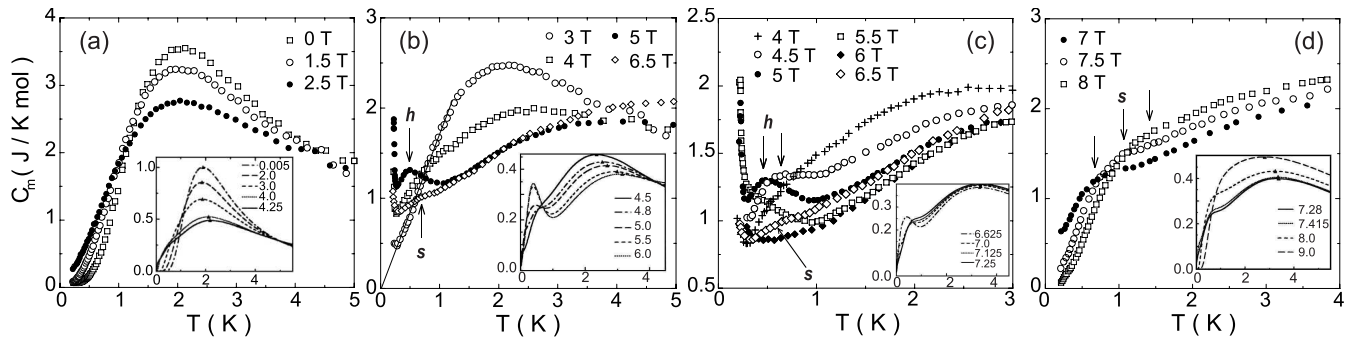}
\caption{\label{fig:gapless} Temperature dependence of $C_\mathrm{m}$ in magnetic fields. 
(a) 0 T$\leq H \leq$2.5 T (spin-gapped phase)
(b) and (c) 3 T$\leq H \leq$6.5 T (gapless phase)
(d) 7 T$\leq H \leq$8 T (spin-polarized phase).
A solid line in (b) shows linear-$T$ dependence of $C_\mathrm{m}$. 
Down-arrows in (b), (c) indicate a hump in $C_\mathrm{m}$. 
Up-arrows in (b), (c) and down-arrows in (d) indicate a shoulder $C_\mathrm{m}$. 
Insets show theoretical calculations of the strongly coupled two-leg ladder system 
\cite{wang}. }
\end{figure*}

In Fig. 3(a), we show $C_\mathrm{m}$ in the field range of $H \leq$ 2.5 T. 
All data exhibit the exponential decay characteristic of the spin-gapped phase at low temperatures. 
With increasing magnetic field, the exponential behavior gradually changes to the almost linear-$T$ dependence, 
indicating the suppression of the energy gap. 
It is noteworthy that $C_\mathrm{m}$ does not seem to approach zero monotonously as $T$ $\rightarrow$ 0 K at $H=2.5$ T, 
which may imply a certain structure in $C_\mathrm{m}$ at lower temperatures. 
Because 2.5 T is lower than $H_\mathrm{C1}$, 
it is hard to consider that this behavior is relevant to the FIMO.
We will discuss this point later.

Figure 3(b) shows the results of $C_\mathrm{m}$ in the field range of 3.0 T $ \leq H \leq$ 6.5 T, 
which corresponds to the gapless phase 
between $H_\mathrm{C1}$ and $H_\mathrm{C2}$ estimated by the magnetization study \cite{Hosokoshi}. 
The characteristic behaviors in this phase are emergence and shift of the hump 
and appearance of the shoulder depending on the field strength outside of the FIMO phase in the $H$-$T$ phase diagram. 
At 3 T, an up-turn deviating from linear-$T$ dependence is seen at lowest temperatures, 
which can be due to the FIMO.  
The up-turn is enhanced with increasing field up to 5 T, and seems to be suppressed in the higher field. 
At $H=5$ T, it is clear that the hump emerges at around 500 mK as indicated by the down-arrow, 
followed by the sharp anomaly due to the FIMO at $T\simeq  200$ mK as mentioned before.    
Further, a shoulder emerges at around $T \simeq $ 700 mK at $H=6.5$ T as indicated by the up-arrow. 
   
For a closer look of the results in this region,  
the low temperature parts of $C_\mathrm{m}$ for 4 T $\leq H \leq$ 6.5 T are shown in Fig. 3(c). 
At $H=4.5$ T, the hump appears at $T \simeq$ 600 mK. 
As increasing field, the hump shifts to the lower temperatures.  
Finally, the hump merges with the sharp anomaly due to the FIMO at $H=5.5$ T 
and is not observed at $H=6$ T. 
From the emergence and shift of the hump, 
it is reasonable to consider that the hump is masked by the broad maximum 
due to the short-range ordering in the field range of 3 T$\leq H\leq 4$ T. 
While the hump already has not been seen at all at 6 T, the shoulder emerges at 6.5 T at around 700 mK 
above the temperature range where the sharp anomaly 
due to FIMO 
is seen.
The origin of the behaviors will be discussed later.

Figure 3(d) shows temperature dependence of $C_\mathrm{m}$ for $H$$\geq$7 T, where the system is in the spin-polarized phase. 
We can see the exponential decay at low temperatures as in the case for $H$$\leq$2.5 T.
In this phase, the spins are forced to align parallel by a strong magnetic field.  
That is, the decay indicates 
an energy gap induced by the strong magnetic field.  
The shoulder in $C_\mathrm{m}$ is also seen in the temperature range of 600 mK$\leq T<1$ K 
as in the data at 6.5 T in Figs. 3(b) and (c). 
The magnitude of this shoulder at 7 T is larger than that at 6.5 T and locates at higher temperatures. 
This shoulder further develops with increasing magnetic field.  
It is noted that $C_\mathrm{m}$ does not seem to approach zero monotonously down to 0 K 
at $H=7$ T implying a certain structure in $C_\mathrm{m}$ at lower temperatures as in the case at $H=2.5$ T. 



To discuss the present results, we first refer to a theoretical study by Tachiki and Yamada about the specific heat for weakly coupled $S=1/2$ 
dimer systems, which include the alternating chain system and the two-leg ladder system 
in certain parameter regions, in magnetic fields \cite{tachiki & yamada}. 
In the study they obtain an effective Hamiltonian in magnetic fields around $H_\mathrm{C1}$ 
by neglecting the upper two levels in the energy spectrum of each dimer and  
replacing the lower two levels, corresponding to the singlet ground state and the lowest branch of the triplet, 
by the effective spin $S=1/2$. 
This replacement allows us to describe the systems by the Hamiltonian of a 1D $S=1/2$ XXZ model with easy-plane anisotropy in the effective field.
It is noted that physical properties of the alternating chain system and the two-leg ladder system 
in magnetic fields are expected to be qualitatively same from this mapping of the Hamiltonian.
Indeed, the specific heat of F$_5$PNN is similar with that of an $S=1/2$ two-leg ladder 
system 
CuHpCl, above the critical temperature of the FIMO [6-8]. 
A hump in $C_\mathrm{m}$ is seen in the gapless phase of both compounds.
Whereas, the NMR relaxation rate $T_1 ^{-1}$ of 
CuHpCl 
shows an anomalous increase with the power-law temperature 
dependence as $T$$\rightarrow$$0$ K, which is characteristic of the TLL, 
in the magnetic field and temperature range where the hump in $C_\mathrm{m}$ was observed \cite{chabou}. 
In the case of F$_5$PNN, the power-law behavior in $T_1 ^{-1}$ is also observed below about 4 K 
in the gapless phase, as mentioned before \cite{izumi}. 
This temperature-field range includes that where the hump in $C_\mathrm{m}$ is seen. 

Then, we refer to a transfer-matrix renormalization group study by Wang {\it et al}. 
for thermodynamic properties of an $S=1/2$ two-leg ladder model in magnetic field 
with parameters $J_{\parallel} =1$ and $J_{\perp} /J_{\parallel} =5.28$, 
where $J_{\parallel}$ and $J_{\perp}$ are the intra- and inter-chain coupling, respectively \cite{wang}.
In this model, the two critical fields are estimated as 
$H_\mathrm{C1}=4.3823$ and $H_\mathrm{C2}=7.2800$. 
Our results are qualitatively in good agreement with their calculations on the whole phases.
Their calculations for $H<H_\mathrm{C1}$, $H_\mathrm{C1}\leq H\leq H_\mathrm{C2}$, and $H>H_\mathrm{C2}$
are shown in the insets of Figs. 3(a), (b) and (c), (d) respectively. 
In the following, we will analyze our results by comparing them to their calculations in each phase. 

In the spin-gapped phase below $H_\mathrm{C1}$ (Fig. 3(a) inset), their calculations show 
a single peak structure due to the short-range ordering and suppression of the energy gap 
by applying magnetic field. 
It should be noted that their results show a cusp just below $H_\mathrm{C1}$ at very low temperatures. 
Turning now to the present experimental results, the data at 2.5 T, where is just below $H_\mathrm{C1}$, do not seem to 
approach zero monotonously down to 0 K as mentioned before. 
From the good qualitative agreement between both results, 
it is considered that the above behavior at 2.5 T is ascribed to the cusp in their calculations. 
Therefore, we will clearly observe the similar cusp if we can make the measurement at lower temperatures possible. 
Wang {\it et al.} suggest that this cusp appears at the vicinity of boundary between the spin-gapped and TLL phase 
in the $H$-$T$ phase diagram obtained from the calculations of the specific heat and the magnetization. 

In the field range of $H_\mathrm{C1}<H<H_\mathrm{C2}$ (the gapless phase : Figs. 3(b) and (c) insets), 
it is the most significant feature in their calculations that a hump emerges and then shifts to low temperatures as 
increasing magnetic field and a shoulder emerges at $H\simeq H_\mathrm{C2}$. 
This feature is also seen in our results as mentioned before. 
We estimated the ratio of the temperature of the hump with its maximum to
that of the broad maximum due to the short-range ordering in the 1D magnetic systems
in both our results and theoretical results. 
From this estimation, the ratio in our results is 
roughly consistent with that in theoretical results. 
(In both cases, the ratios are about 10 \% in the middle field of the gapless
field region.)  
According to Wang {\it et al.}, the hump comes from low-energy excitations in the TLL and 
the linear-$T$ dependence should be observed at low temperatures. 
However, the linear-$T$ dependence in $C_\mathrm{m}$ is not seen in our results 
because it is masked by the anomaly due to FIMO at lower temperatures. 
As for the origin of the shoulder, they do not make a definite remark  
but suggest that it appears at the vicinity of the TLL phase.

In the spin-polarized phase for  $H\geq H_\mathrm{C2}$ (Fig. 3(d) inset),  
the development of the shoulder is clearly seen in their calculations 
as well as our experimental results.
Moreover, another cusp is seen at low temperatures just above $H_\mathrm{C2}$. 
In the corresponding field of our results ($H=7$ T), 
$C_\mathrm{m}$ does not seem to approach monotonously zero as $T$$\rightarrow$0 K. 
From the good agreement between both results, our result at 7 T can correspond to this cusp as in the case at 2.5 T. 
This cusp also locates at the vicinity of boundary of the TLL phase in their $H$-$T$ phase diagram. 

Finally, to evaluate the validity of the above comparison, 
we will show the field dependence of the maximum magnetic specific heat $C^\mathrm{max} _\mathrm{m}$ 
and the corresponding temperature $T_\mathrm{max}$ of the present results in Fig. 4. 
As increasing magnetic field, $C^\mathrm{max} _\mathrm{m}$ first declines and shows a minimum around the field value of ($H_\mathrm{C1}+H_\mathrm{C2})/2$. 
Then, $C^\mathrm{max} _\mathrm{m}$ begins to increase for the higher fields up to 8 T. 
On the other hand, $T_\mathrm{max}$ keeps its original value and 
begins to increase at $H\simeq H_\mathrm{C1}$ and reaches a maximum around $H_\mathrm{C2}$. 
These features of field dependence of $C^\mathrm{max} _\mathrm{m}$ and $T_\mathrm{max}$ 
well agree with the theoretical results by Wang {\it et al.} (Fig. 6 of Ref. 12).
A theoretical calculation is required for the present alternating Heisenberg chain system with the experimental features 
mentioned above.


In conclusion, we have performed the detailed specific heat measurement 
on the $S=$1/2 alternating Heisenberg chain system F$_5$PNN 
in magnetic fields up to 8 T.
This is the first detailed experiment on the specific heat 
of the spin-gapped system up to above $H_\mathrm{C2}$.
The characteristic behaviors are observed at around the gapless field region outside of the FIMO phase, 
which well agree with the theoretical calculation 
in the TLL regime qualitatively \cite{wang}. 
Thus, this agreement suggests that the observed behaviors are related with the low-energy exicitation in TLL state. 
This scenario is also supported by the power-law behavior in the NMR relaxation rate $T_1 ^{-1}$ of F$_5$PNN observed 
below 4 K in the gapless field region \cite{izumi}. 



We would like to express our sincere thanks to A. Tanaka, K. Izumi, T. Goto, 
S. Suga, N. Maeshima and T. Sakai for valuable discussions. 

\begin{figure}[htbp]
\includegraphics*[width=5cm]{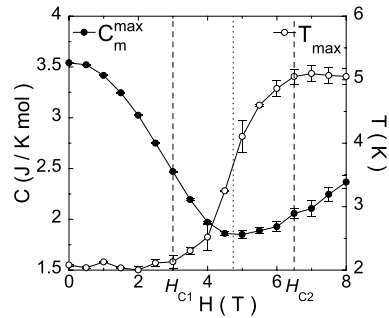}
\caption{\label{fig:FI-gapped} The maximum magnetic specific heat $C^\mathrm{max} _\mathrm{m}$ and the corresponding temperature 
$T_\mathrm{max}$ versus magnetic field $H$. Errors at around $H_\mathrm{C2}$ are due to the broader maximum of $C_\mathrm{mag}$.}
\end{figure}


\end{document}